\documentclass[letterpaper, 10 pt, conference]{ieeeconf} 
\pdfoutput=1 
\IEEEoverridecommandlockouts
\usepackage[english]{babel}
\usepackage{cite}
\usepackage{amsmath,amssymb,amsfonts}
\usepackage{algorithm}
\usepackage{algorithmic}
\usepackage{graphicx}
\usepackage{textcomp}
\usepackage{amsmath}
\usepackage{dsfont}
\usepackage{xcolor}
\usepackage{units}
\usepackage{booktabs}
\usepackage{comment}
\usepackage{url}
\usepackage[normalem]{ulem}
\usepackage[hidelinks]{hyperref}
\usepackage{caption}
\usepackage{subcaption}
\usepackage{flushend}
\usepackage{lettrine}

\usepackage{amsthm}
\newcounter{thm}
\newtheorem{prob}[thm]{Problem}
\newtheorem{lemma}{Lemma}[section]
\newtheorem{theorem}{Theorem}[section]
\newtheorem{definition}{Definition}[section]
\newtheorem{remark}{Remark}[section]

\def\BibTeX{{\rm B\kern-.05em{\sc i\kern-.025em b}\kern-.08em
    T\kern-.1667em\lower.7ex\hbox{E}\kern-.125emX}}
\newcommand{\flexbrac}[1]{\if\relax\detokenize{#1}\relax \else (#1) \fi}
\newcommand{\flexcomma}[1]{\if\relax\detokenize{#1}\relax \else ,#1 \fi}

\newcommand{\abs}[1]{\lvert #1 \rvert}

\newcommand{\cA}{\mathcal{A}}

\newcommand{\cG}{\mathcal{G}}

\newcommand{\cR}{\mathcal{R}}

\newcommand{\cV}{\mathcal{V}}

\newif\ifmargincomments 
\margincommentstrue

\ifmargincomments

\else

\fi

\begin{document}

\title{\LARGE \bf Congestion-aware Ride-pooling in Mixed Traffic \\ for Autonomous Mobility-on-Demand Systems}

\author{Fabio Paparella, Leonardo Pedroso, Theo Hofman, Mauro Salazar
\thanks{Control Systems Techonology section, Eindhoven University of Technology, Eindhoven, The Netherlands
        {\tt\small \{f.paparella, l.pedroso,t.hofman,m.r.u.salazar\}@tue.nl}}%
}


\maketitle
\thispagestyle{empty}
\pagestyle{empty}



\begin{abstract}
This paper presents a modeling and optimization framework to study congestion-aware ride-pooling Autonomous Mobility-on-Demand (AMoD) systems, whereby self-driving robotaxis are providing on-demand mobility, and users headed in the same direction share the same vehicle for part of their journey.
Specifically, taking a mesoscopic time-invariant perspective and on the assumption of a large number of travel requests, we first cast the joint ride-pooling assignment and routing problem as a quadratic program that does not scale with the number of demands and can be solved with off-the-shelf convex solvers.
Second, we compare the proposed approach with a significantly simpler decoupled formulation, whereby only the routing is performed in a congestion-aware fashion, whilst the ride-pooling assignment part is congestion-unaware.
A case study of Sioux Falls reveals that such a simplification does not significantly alter the solution and that the decisive factor is indeed the congestion-aware routing.
Finally, we solve the latter problem accounting for the presence of user-centered private vehicle users in a case study of Manhattan, NYC, characterizing the performance of the car-network as a function of AMoD penetration rate and percentage of pooled rides within it.
Our results show that AMoD can significantly reduce congestion and travel times, but only if at least 40\% of the users are willing to be pooled together. Otherwise, for higher AMoD penetration rates and low percentage of pooled rides, the effect of the additional rebalancing empty-vehicle trips can be even more detrimental than the benefits stemming from a centralized routing, worsening congestion and leading to an up to 15\% higher average travel time.
\end{abstract}



\section{Introduction}
\begin{figure}[t]
	\centering
	\begin{subfigure}{\linewidth}
		\centering
		\includegraphics[width=.75\linewidth]{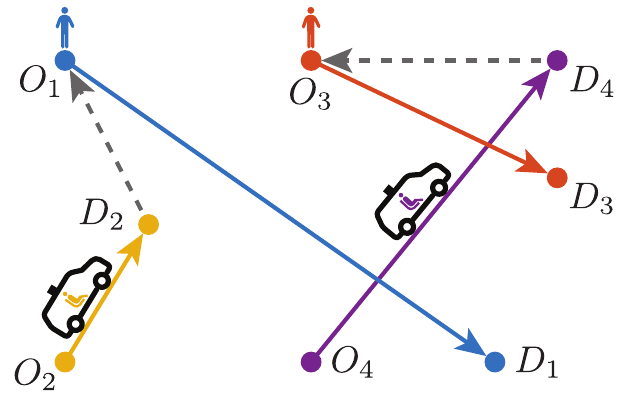}
		\caption{Individual ride-sharing.}
	\end{subfigure}\\
	\vspace{0.5cm}
	\begin{subfigure}{\linewidth}
		\centering
		\includegraphics[width=.75\linewidth]{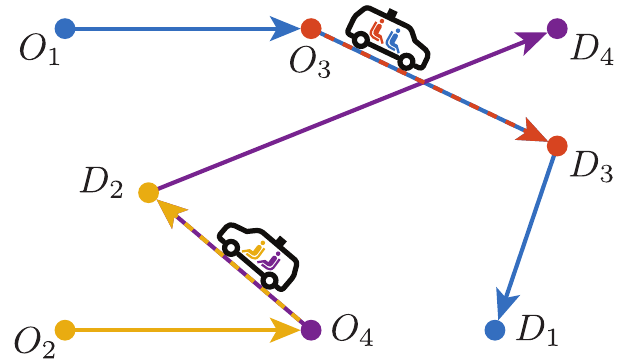}
		\caption{Ride-pooling.}
	\end{subfigure}
	\caption[]{Illustrative comparison of individual ride-sharing and ride-pooling with two vehicles; dashed lines represent rebalancing flows without users aboard.\footnotemark}
	\label{fig:eg_ridepooling}
	\vspace{-0.5cm}
\end{figure}

\lettrine{R}{ide-sharing} is the concept of passengers sharing a common vehicle throughout a day.
Withing this scheme, ride-pooling paradigms pool multiple passengers, heading in the same direction, together, and transport them in the same vehicle as depicted in Fig.~\ref{fig:eg_ridepooling}, resulting into lower costs, reduced emissions and reduced vehicle hours traveled.
To provide such mobility-as-a-service in an on-demand fashion, it is possible to leverage Autonomous Mobility-on-Demand (AMoD) systems, fleets of autonomous vehicles that provide on-demand mobility and can be centrally coordinated. 
Understanding how to pool together requests, the so-called ride-pooling assignment, is a notoriously difficult combinatorial task~\cite{Santi2014,Alonso_Mora_2017}. On top of that, albeit more holistic, taking into account endogenous and exogenous congestion patterns in the assignment problem further increases the complexity of the problem. 
In this paper, we study two-person ride-pooling from a stochastic point of view in a time-invariant multi-commodity network flow model, a mesoscopic modeling approach usually used for mobility planning and design. Specifically, we address the ride-pooling assignment problem from a rebalancing- and congestion-aware perspective. In addition, we model the presence of private users that cannot be controlled by the AMoD operator and react to congestion selfishly to minimize their own travel time.

\footnotetext{Icon ``car" of package Lucide is used in Fig.~\ref{fig:eg_ridepooling} under the ISC License~\cite{ISC_license}: Copyright (c) for portions of Lucide are held by Cole Bemis 2013-2022 as part of Feather (MIT). All other copyright (c) for Lucide are held by Lucide Contributors 2022. Icon ``seat-passenger" of package Material Design Icons is used in Fig.~\ref{fig:eg_ridepooling} under the Apache 2.0 License \cite{Apache_license}: Copyright 2023 Pictogrammers.}

\emph{Related Literature:}
Ride-pooling has been extensively studied in recent years~\cite{TsaoMilojevicEtAl2019}.
In~\cite{Santi2014}, the authors analyzed the impact and benefits of ride-pooling. Their study in Manhattan shows that approximately 80\% of the taxi trips can potentially be shared by two requests, at a small inconvenience of a few minutes of additional travel time. However, the methods presented therein are limited to two people with optimality guarantees and three people resorting to heuristics. In addition, these solutions are intractable for larger number of passengers. To bridge the gap, Alonso-Mora et al.~\cite{Alonso_Mora_2017} developed an efficient algorithm to compute optimal routing for real-time applications in a ride-pooling environment with high-capacity vehicles. In~\cite{FielbaumKucharskiEtAl2022}, the authors explored how to split costs between requests pooled together, while in~\cite{FielbaumTirachiniEtAl2021} the same authors analyzed the performance of such ride-pooling systems as a function of its scale. However, every aforementioned solution is based on a microscopic model, whereby the complexity of the resulting problem grows with the number of travel requests. 

Mesoscopic models for AMoD range from fluidic methods~\cite{PavoneSmithEtAl2012} to queuing theoretical models~\cite{ZhangPavone2014}. They enable an efficient inclusion of a wide variety of constraints and joint optimization with connected systems~\cite{
ZardiniLanzettiEtAl2022,Rossi2018,RossiIglesiasEtAl2018b}, but they do not allow to seamlessly account for ride-pooling.
Recently, we proposed a methodology to capture ride-pooling within time-invariant network flow models~\cite{PaparellaPedrosoEtAl2023}, whereby the ride-pooling request assignment was carried out neglecting rebalancing flows, as well as endogenous and exogenous congestion effects caused by the AMoD fleet and private vehicles, respectively.
To conclude, to the best of the authors' knowledge, a rebalancing and congestion-aware ride-pooling mesoscopic framework in mixed traffic has not been proposed to date.

\emph{Statement of Contributions:}
The contribution of this work is fourfold: i)~We propose a 
congestion-aware routing two-person ride-pooling time-invariant network flow model in mixed traffic; ii)~we jointly formulate the congestion-aware routing problem, the rebalancing- and congestion-aware ride-pooling assignment problem, and cast them as a quadratic program~(QP), whose dimension does not grow with the number of travel requests; iii) leveraging  simulations in the Sioux Falls network, we study the performance improvement that stems from a rebalancing- and congestion-aware assignment against an unaware one, but where the routing is still congestion-aware, and quantify the benefits of ride-pooling as a function of number of requests per unit time and penetration rate compared to private users; and iv) we perform simulations in Manhattan, and quantify and analyze the impact of AMoD and ride-pooling penetration rates under heavy congestion and mixed traffic settings.


\emph{Organization:} The remainder of this paper is organized as follows: Section~\ref{sec:model} presents the multi-commodity traffic flow problem employed to capture congestion-aware routing of ride-pooling in mixed-traffic with congestion-unaware assignment.
In Section~\ref{sec:assignment}, the rebalancing- and congestion-aware ride-pooling assignment and routing problem is cast as a quadratic program. In Section~\ref{sec:res}, we illustrate the proposed approach in case-studies of Sioux Falls and Manhattan. Finally, in Section~\ref{sec:conc}, we outline the major conclusions of this work. 

\emph{Notation:}  We denote the vector of ones, of appropriate dimensions, by $\mathds{1}$. The $i$th component of a vector $v$ is denoted by $v_i$ and the entry $(i,j)$ of a matrix $A$ is denoted by $A_{ij}$. The cardinality of set $\mathcal{S}$ is denoted by $\abs{\mathcal{S}}$.

\section{Ride-pooling Network Flow Model}\label{sec:model}

In this section, first, we briefly introduce the network flow model proposed in \cite{PavoneSmithEtAl2012} in mixed traffic conditions, similarly to \cite{Wollenstein-BetechSalazarEtAl2021}. Then, we reformulate the problem to capture ride-pooling. Last, we provide a brief discussion of the model.

\subsection{Time-invariant Network Flow Model} 
 The transportation network is modeled a directed graph $\mathcal{G} = (\mathcal{V}, \mathcal{A})$.
 The set of vertices $\mathcal{V} := \{1,2,...,\abs{\mathcal{V}}\}$ represent intersections, while the set of arcs $\mathcal{A} \subseteq \mathcal{V} \times \mathcal{V}$, are road links. The incidence matrix of the network $\mathcal{G}$ is denoted by ${B \in \{-1,0,1\}^{\abs{\cV} \times \abs{\cA}}}$~\cite{Bullo2018}. 
The travel requests are defined as follows:
\begin{definition}[Travel Requests]\label{def:requests}
	A travel request is a tuple $r = (o,d,\alpha) \in \mathcal{V} \times \mathcal{V} \times \mathbb{R}_{>0}$, where $\alpha$ is the number of users per unit time traveling from $o$ to $d \neq o$.  The set of requests is $\mathcal{R} := \{r_m\}_{m\in \mathcal{M}}$, where $\mathcal{M} = \{1,\ldots,M\}$.
\end{definition} 
In this paper, we define active vehicle flows as the flows of vehicles with users on-board, while the rebalancing flows correspond to the empty vehicle flows required to realign the vehicles from drop-off to the pick-up locations. 
We model the flows of active vehicles as a matrix $X \in \mathbb{R}^{\abs{\cA} \times \abs{\cV}}$ defined as $X := \left[x^{1}\  x^2 \, \dots \,x^{\abs{\cV}}\right]$, where  $x^{i} \in \mathbb{R}^{\abs{\cA}}$ is the vector of vehicle flows that share the same origin $i \in \cV$. Similarly, the rebalancing flow is denoted by $x^{\mathrm{r}} \in \mathbb{R}^{\abs{\cA}}$, while $x^{\mathrm{p}} \in \mathbb{R}^{\abs{\cA}}$ corresponds to the exogenous flow of private vehicles. The travel time vector $t(X,x^r,x^\mathrm{p}) \in \mathbb{R}^{\abs{\cA}}$ 
is a function of the active vehicle flows, rebalancing flows, and private user flows on each arc.  
We account for mixed-traffic conditions. Let  $\phi\in [0,1]$ denote the penetration rate of the AMoD fleet, i.e., the fraction of AMoD users compared to the total amount of users. Similarly, to the AMoD travel requests defined in Definition~\ref{def:requests}, one may analogously define a set of private exogenous requests $\mathcal{R}^\mathrm{p}$. We model the private users' decision as a user-centric  traffic assignment problem (TAP) routing, whereby each private user minimizes their own travel time given certain traffic conditions \cite{Patriksson2015}. Indeed, given private exogenous requests $\mathcal{R}^p$, active flows $X$, and rebalancing flows $x^\mathrm{r}$, we denote the solution of resulting user-centric flows as $x^\mathrm{p} \in \mathrm{TAP}(\mathcal{R}^\mathrm{p}, X,x^\mathrm{r})$.  A detailed analysis on the TAP and on the methods that may be employed to solve it jointly with time-invariant network flow models has been addressed in~\cite{Wollenstein-BetechSalazarEtAl2021}. In the following, similarly to \cite{Wollenstein-BetechSalazarEtAl2021}, we define the mixed-traffic bilevel network flow problem.


\begin{prob}[Multi-commodity Network Flow Problem]\label{prob:main}
Given a road graph $\cG$ and a set of travel requests $\cR$, the optimal active vehicle flow $X$ and rebalancing flow $x^\mathrm{r}$ result from
	\begin{equation*}
		\begin{aligned}
			\min_{X, x^\mathrm{r}}\; &J(X,x^\mathrm{r}) =t(X,x^r,x^\mathrm{p})^\top ( X \mathds{1} + \rho x^\mathrm{r} )  \\
			\mathrm{s.t. }\; & BX = D \\
			&B ( X \mathds{1}+ x^\mathrm{r} )=0 \\
			& X, x^\mathrm{r} \geq 0\\
			&x^\mathrm{p} \in \mathrm{TAP}(\mathcal{R}^\mathrm{p}, X,x^\mathrm{r}),
		\end{aligned}
	\end{equation*}
	where $\rho \in [0,1]$ is a parameter, $D \in \mathbb{R}^{\abs{\mathcal{V}} \times \abs{\mathcal{V}}}$ is the travel requests demand matrix between each pair of vertices, whose entries are 
	\begin{equation}\label{eq:def_D}
		\!\!\!\!D_{ij} = \begin{cases}
			\alpha_m, &  \exists m \in \mathcal{M} : o_m = j \land d_m = i\\
			-\sum_{k\neq j} D_{kj}, & i  = j \\ 
			0, &   \mathrm{otherwise},
		\end{cases}\!\!\!
	\end{equation}
\end{prob}
For $\rho =1$, the objective function of Problem~\ref{prob:main} represents the total vehicle travel time of the fleet~\cite{PavoneSmithEtAl2012,Rossi2018}. For $\rho =0$, the objective function is the total user travel time.

\subsection{Ride-pooling Time-invariant Network Flow Model}\label{sec:ride-pool-model}

Recently, in \cite{PaparellaPedrosoEtAl2023}, we proposed a formulation to take into account ride-pooling in network flow models by transforming the original set of requests, $D$, into an equivalent set that encodes ride-pooling trips, $D^\mathrm{rp}$. Analogously, we extend the formulation therein to define the ride-pooling network flow problem in mixed-traffic conditions as follows:

\begin{prob}[Ride-pooling Network Flow Problem]\label{prob:rides}
	Given a road graph $\cG$ and a ride-pooling assignment matrix $D^\mathrm{rp}$, the optimal active vehicle flow $X$ and rebalancing flow $x^\mathrm{r}$ result from
	\begin{equation*}
		\begin{aligned}
			\min_{X,x^\mathrm{r}}\; &J(X,x^\mathrm{r}) = {t(X,x^\mathrm{r},x^\mathrm{p})^\top (X \mathds{1} + \rho x^\mathrm{r})} \\
			\mathrm{s.t. }\; & BX = D^\mathrm{rp} \\
			&B ( X \mathds{1}+ x^\mathrm{r} )= 0 \\
			& X, x^\mathrm{r} \geq 0\\
			&x^\mathrm{p} \in \mathrm{TAP}(\mathcal{R}^\mathrm{p}, X,x^\mathrm{r}).
		\end{aligned}
	\end{equation*}
\end{prob}

The ride-pooling assignment matrix $D^\mathrm{rp}$ in Problem~\ref{prob:rides} describes the ride-pooling assignment pattern, i.e., how the different requests are matched to be ride-pooled together. Four key conditions must be met when designing $D^\mathrm{rp}$, which are thoroughly detailed and analyzed in \cite{PaparellaPedrosoEtAl2023} and that we also systematically address here: i)~All the individual requests, described by $D$, must be served, whether they are ride-pooled or not; ii)~ride-pooling two requests is feasible from a spatial point of view only if the detour time (additional travel time of ride-pooling compared to serving the request individually) of both requests is below a threshold $\bar{\delta} \in \mathbb{R}_{ \geq 0}$; iii)~ride-pooling two requests is feasible from a temporal point of view only if the waiting time for a request to start being served is below a threshold $\bar{t} \in \mathbb{R}_{>0}$; and iv)~the requests are assigned to be ride-pooled in $D^\mathrm{rp}$ such that the objective function of Problem~\ref{prob:rides} at its solution is minimized.

\subsection{Discussion on the Congestion-unaware Assignment}

In \cite{PaparellaPedrosoEtAl2023}, we proposed a knapsack-like approach to optimally compute the ride-pooling demand matrix $D^\mathrm{rp}$ in a rebalancing-unaware, i.e, with ${\rho = 0}$, and congestion-unaware setting, i.e., assuming the arc travel times are constant. Then, with that ride-pooling assignment matrix, Problem~\ref{prob:rides} can be solved for the AMoD flows for any $\rho \in [0,1]$. Notwithstanding the combinatorial nature of matching travel requests, this approach, supported by the network flow formulation, allowed to attain a computationally feasible polynomial-time solution. The natural next step is to reflect on the role that congestion and rebalancing play in the design of the ride-pooling demand matrix $D^\mathrm{rp}$. However, it is not possible to extend this approach to a rebalancing- or congestion-aware setting. Indeed, the computation of the ride-pooling matrix in Problem~\ref{prob:rides} is decoupled from the flow optimization. As a result, it does not allow to take into account the effect of the ride-pooling allocation on the travel-times, when computing matrix $ D^\mathrm{rp}$. This observation motivates the need to jointly optimize for the ride-pooling assignment and vehicle flows in a rebalancing- and congestion-aware setting, as shown in Section~\ref{sec:assignment} below.

\subsection{Discussion of the Model}

A few comments are in order regarding the validity of the model. First, the mobility system is analyzed at steady-state in a time-invariant framework. As a consequence, this model has been mainly used for planning and design purposes by several works in the literature. This assumption is reasonable if the travel requests vary slowly w.r.t. the average time of serving each request. This is the case especially in highly populated metropolitan areas \cite{IglesiasRossiEtAl2017}. Second, our framework does not take into account the stochastic nature of the exogenous congestion that determines the travel time in each road arc. However, this deterministic approach is suitable for our purposes as it provides an average representation of these stochastic phenomena in a mesoscopic scale \cite{Neuburger1971}. Finally, the flows of Problems~\ref{prob:main} and \ref{prob:rides} are continuous, which is in line with a mesoscopic perspective~\cite{PaparellaChauhanEtAl2023,SalazarLanzettiEtAl2019}. 

\section{Congestion-aware Ride-pooling Assignment}\label{sec:assignment}
In this section,  we build upon Problem~\ref{prob:rides} as a means of including the joint optimization of the ride-pooling assignment in the rebalancing- and congestion-aware routing problem.

\subsection{Framework}\label{sec:apprx}

We assume that the travel time function in each arc, $t_a(X,x^r,\!x^\mathrm{p})$, is strictly increasing w.r.t.\ $X_{ai}, \,i\in \mathcal{V}$, $x^r_a$, and $x^\mathrm{p}_a$. We denote the free-flow travel time of link $a$ by  $t_a^0$. No further assumptions are made, making the development of the method generic.  Afterwards, we particularize the approach for the Bureau of Public Roads (BPR) function~\cite{BPR1964}, whose piece-wise linear approximation can be leveraged to cast the joint optimization as a QP optimization problem~\cite{Wollenstein-BetechSalazarEtAl2021}. The BPR function is expressed as 
\begin{equation}\label{eq:BPR}
	t_a(X,x^r,\!x^\mathrm{p}) = t_a^0\left(\!1\!+\!\alpha\left(\frac{\sum_{i\in \mathcal{V}} X_{ai} + x^r_a + x^\mathrm{p}_a}{\kappa_a}\right)^{\!\!\beta}\right)\!,
\end{equation}
where $\kappa_a$ is the capacity of link $a$, and $\alpha, \beta$ are two constants, typically $\alpha = 0.15$ and $\beta = 4$.
We  also assume high demand conditions, which is in order for highly populated metropolitan areas and is quantified in what follows.

We now formulate the ride-pooling assignment problem that stems from the four conditions outlined in Section~\ref{sec:ride-pool-model}. In the following subsections, we introduce the approximations that we leverage to approximate it as a QP.

\subsubsection{Spatial Analysis}\label{sec:SD}
We analyze how to enforce feasibility of ride-pooling between two requests from a spatial perspective in a congestion-aware setting. If the detour time experienced by any of the two users is higher than the threshold $\bar{\delta}$, ride-pooling those two requests is unfeasible. However, in a congestion-aware setting, the detour time depends on the ride-pooling assignment, because it defines the flows of vehicles in the network which, in turn, define the travel times in the arcs. Although it is possible to formulate this constraint, it cannot be expressed linearly. As a result, we introduce an approximation. The main premise is that the travel times, and consequently detour times, in a congested setting are greater than in free-flow conditions. As a result, if ride-pooling two requests is unfeasible in free-flow conditions, then it is also unfeasible in congested conditions. On top of that, since the flows and ride-pooling assignment are optimized jointly to minimize travel time, the configurations with lower detours are prioritized. Thus, our proposed approach is to only consider ride-pooling assignments that are feasible in free-flow conditions and then optimize the user demand that is assigned to each of them. Nevertheless, there is no guarantee that the original threshold constraint will not be violated.



\begin{figure}[t]
	\centering
	\includegraphics[width = 0.65\linewidth]{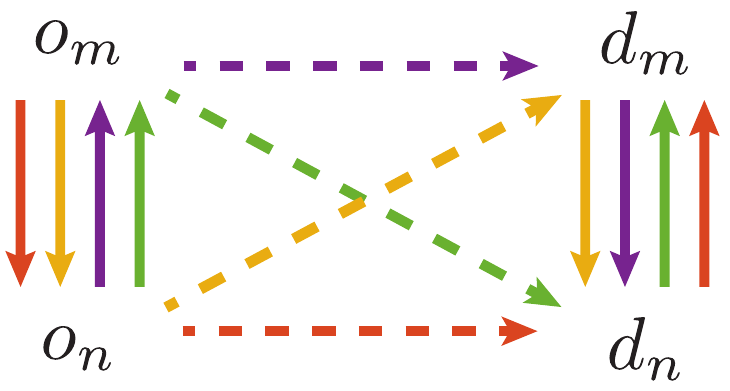}
	\caption{Four configurations to ride-pool two requests $r_m,r_n \in \cR$. The dashed arrows represent a vehicle flow with two users, whilst the solid ones represent a flow with a single user. Taken from~\cite{PaparellaPedrosoEtAl2023}.}
	\label{fig:serve_conf}
	\vspace{-0.5 cm}
\end{figure}

Consider two requests $r_m,r_n\in \cR$. The goal is to assess whether it is feasible or not to ride-pool $r_m$ and $r_n$ in free-flow conditions. There are four different ways of serving two requests $r_m, r_n \in \cR$, as depicted in Fig.~\ref{fig:serve_conf}. Each configuration can be split into three equivalent vehicle travel requests, each corresponding to an arrow in the figure. For example, the green configuration in Fig.~\ref{fig:serve_conf}, whose sequence of visited nodes is $(o_n,o_m,d_n,d_m)$, can be split into the vehicle flow requests $(o_n,o_m)$, $(o_m,d_n)$, and $(d_n,d_m)$. Denote the set of such equivalent requests for configuration $c$ as $\cR_{mn}^c$ ($\cR_{nm}^c = \cR_{mn}^c$) and we define $\Pi(\cR_{mn}^c)$ as the sequence of visited nodes. Let us define the set of configurations that are feasible to be ride-pooled in free-flow conditions as $\mathcal{P}_{mn}$. For each configuration $c$, one can now solve Problem~\ref{prob:rides}, for $\rho = 0$,  with a simplified demand matrix $D^\mathrm{rp} = D^{mn,c}$ obtained from the set of requests $\cR_{mn}^c$ with \eqref{eq:def_D}, obtaining a flow $X^{mn,c} \in \mathbb{R}^{\abs{\mathcal{V}}\times \abs{\mathcal{V}}}$. The delay of request $r_m$ for a configuration $c$ is then given by
\begin{equation*}
	\delta^{mn,c}_m = \sum_{p \in {\pi^c_{mn}} } [{t^0}^\top X^{mn,c}]_p   - [{t^0}^\top X^{mn,0}]_{o_m},
\end{equation*}
where ${\pi^c_{mn}} \subseteq \Pi(\cR_{mn}^c)$ is the ordered set of nodes in $\Pi(\cR_{mn}^c)$ from $o_m$ to the node before $d_m$, ${t^0}^\top:= [t^0_1 \;\cdots \; t^0_{|\mathcal{V}|}]^\top$ is the free-flow travel time vector of the arcs, and $X^{mn,0}$ is the flow matrix that results from serving $r_m$ and $r_n$ individually in free-flow conditions.
A ride-pooling configuration is feasible if the users' delay is below $\bar{\delta}$. Therefore, 
\begin{equation*}
	\mathcal{P}_{mn}:=  \{c\in \left\{1,\ldots,4\}: \delta^{mn,c}_m \leq \bar{\delta} \land \delta^{mn,c}_n \leq \bar{\delta} \right\}.
\end{equation*}
Note that, if $c \notin \mathcal{P}_{mn}$, then configuration $c$ is not feasible for ride-pooling $r_m$ and $r_n$  under free-flow conditions. It is also important to remark that if $m = n$, then all configurations are feasible, all corresponding to ride-pooling request $r_m$ with itself according to the ordered node visit $(o_m,o_m,d_m,d_m)$.


To conclude, every feasible ride-pooling configuration in free-flow conditions, i.e., $c \in \mathcal{P}_{mn}$, can be a potential candidate for the  ride-pooling assignment.  Henceforth, the simplified demand matrix of each feasible configuration for ride-pooling $r_m$ and $r_n$ is denoted by $D^{mn,c}$. 

%

\subsubsection{Temporal Analysis}\label{sec:TD}
In this section, we analyze the feasibility of ride-pooling two requests from a temporal dimension. Although one can take into account the \mbox{probability} of two requests taking place within the maximum waiting time $\bar{t}$ and assigning the ride-pooling flows taking this into account, it would lead to a combinatorial and nonlinear formulation. For this reason, again, we introduce an approximation leveraging the assumption of high demand.  Following the original idea behind traffic flow models~\cite{PavoneSmithEtAl2012}, the arrival rate of a request $r_m \in \cR$ is described by a Poisson process with parameter $\alpha_m$. Consider two requests $r_m, r_n \in \cR$. First, observe the following result from \cite{PaparellaPedrosoEtAl2023}, which states the probability of two events happening within a maximum time window $\bar{t}$.

\begin{lemma}[{Lemma~II.1 in \cite{PaparellaPedrosoEtAl2023}}]\label{lemma:finalprob}
	Let $r_m, r_n \in \cR$ be two requests whose arrival rate follows a Poisson process with parameters $\alpha_m$ and $\alpha_n$, respectively. The probability of each having an occurrence within a maximum time interval $\bar{t}$ is
	\begin{equation}\label{eq:lem}
		P(\alpha_m,\alpha_n)  := 1-\frac{\alpha_m e^{-\alpha_n\bar{t}}+\alpha_n e^{-\alpha_m\bar{t}}}{\alpha_m+\alpha_n}.
	\end{equation}
\end{lemma}


The key premise of the approximation is that in high demand conditions virtually all travel requests can be ride-pooled. Indeed, the results of \cite{PaparellaPedrosoEtAl2023, Santi2014} support this hypothesis, since they show that, for a large number of travel requests, more than 95\% can be ride-pooled with waiting times lower than \unit[5]{min}. This percentage increases even further for higher waiting times. This is reflected in the approximation that $P(\alpha_m,\alpha_n) \approx 1$, which is reasonable for large $\alpha_m,\alpha_n$. For example, by substituting $\alpha_m,\alpha_n = \unit[15]{requests/h}$ and $\bar{t}=\unit[15]{min}$ in~\eqref{eq:lem} yields $P(\alpha_m,\alpha_n) \approx 0.98$. Thus, in a congested setting with a large number of travel requests and reasonable waiting time, this approximation seems plausible.  Nevertheless, the set of travel requests can be downsized accordingly by a fixed factor, as a means of taking into account that not all the requests can actually be ride-pooled. 


\subsubsection{Expected Number of Pooled Rides}
\label{sec:STF}
In Section~\ref{sec:SD}, we analyzed the spatial dimension of the congestion-aware ride-pooling problem and in Section~\ref{sec:TD} the temporal dimension. By combining these considerations, we formulate the joint ride-pooling time-invariant optimization problem. A fraction of the demand of every request $r_m\in \cR$ can be assigned to be pooled with a request $r_n \in \cR$ if they are spatially feasible in free-flow conditions, i.e., if there is a configuration $c$ whereby $c \in \mathcal{P}_{mn}$. Denote the vehicle demand in pooling requests $r_m,r_n\in \cR$, for a given configuration $c \in \mathcal{P}_{mn}$ as $\gamma_{mn}^c = \gamma_{nm}^c $. The full demand of every $r_m$ is ride-pooled (even if $r_m$ is only pooled with itself) according to the  approximation introduced in Section~\ref{sec:TD}. Thus,
\begin{equation}\label{eq:equality_demands}
	\sum_{n\in \mathcal{M}\setminus \{m\}}\; \sum_{c\in \mathcal{P}_{mn}} \gamma_{mn}^c + 2  \sum_{c\in \mathcal{P}_{mm}} \gamma_{mm}^c  = \alpha_m.
\end{equation}
Note that the factor $2$ is featured in the second term above since, for a given vehicle flow that pools $r_m$ with itself, there is double the user flow that is being served from request $r_m$. Furthermore, whenever a request $r_m$ is pooled with itself, all the configurations correspond to the same route. Thus, we consider a single configuration whose vehicle demand is denoted by $\gamma_{mm}$ and simplified demand matrix by $D^{mm}$. As a result, according to the spatial analysis in Section~\ref{sec:SD}, this pooled vehicle demand is portrayed by the demand matrix $\gamma_{mn}^cD^{mn,c}$ if $m\neq n$ and $\gamma_{mm}D^{mm}$ otherwise.

From the analysis in Section~\ref{sec:TD}, the maximum pooled user demand between $r_m,r_n\in \cR$ is  $\min(\alpha_{m},\alpha_{n})$ if $r_m \neq r_n$ and $\alpha_m/2$ otherwise. As a result, the ride-pooling vehicle flows are bounded as 
\begin{equation*}
	0 \leq \gamma_{mn}^c  \leq \min(\alpha_{m},\alpha_{n}),
\end{equation*}
 $\forall m,n\in\mathcal{M} \;\forall c\in \mathcal{P}_{mn}: m\neq n$, and
\begin{equation*}
	0 \leq \gamma_{mm}  \leq \alpha_{m}/2,\quad \forall m\in\mathcal{M}.
\end{equation*}
Note, however, that if \eqref{eq:equality_demands} is satisfied, then the constraints above are also necessarily satisfied. Therefore, they need not be enforced in the joint optimization problem.

%

Finally, the full ride-pooling demand matrix $D^\mathrm{rp}$ is composed of all ride-pooled active vehicle flows. Thus, the overall ride-pooling assignment matrix can be written as
\begin{equation*}
	D^{\mathrm{rp}} = \sum_{m\in \mathcal{M}} \left(\sum_{\substack{n\in \mathcal{M}\\n>m}}\sum_{c\in \mathcal{P}_{mn}}\gamma_{mn}^c D^{mn,c}+ \gamma_{mm}D^{mm}\right).
\end{equation*}

\subsection{Joint Optimization Approach}

Problem~\ref{prob:rides} can now, finally, be coupled with the ride-pooling assignment matrix $D^\mathrm{rp}$ as follows:

%
%

\begin{prob}[Joint Congestion-aware Ride-pooling Assignment Network Flow Model]\label{prob:ridescongestion}
Given a road graph $\cG$ and a set of travel requests $\cR$, the optimal ride-pooling allocation flows $\gamma_{mn}^c, m,n\in \mathcal{M}, c\in \mathcal{P}_{mn}$, active vehicle flow $X$, and rebalancing flow $x^\mathrm{r}$ result from
	\begin{equation*}
		\begin{aligned}
			&\!\!\!\!\!\!\!\!\!\!\!\!\!\!\!\!\!\!\!\!\!\min_{\substack{X;x^\mathrm{r}; \gamma_{mn}^c,\\ m,n\in \mathcal{M},  n > m, c\in \mathcal{P}_{mn};\\ \gamma_{mm}, m\in \mathcal{M}}}\;  {t(X,x^\mathrm{r},x^\mathrm{p})^\top (X\mathds{1} + \rho x^\mathrm{r})} \\
			\mathrm{s.t. }\; & BX = D^\mathrm{rp}(\gamma) \\
			&B ( X \mathds{1}+ x^\mathrm{r} )= 0 \\
			& X, x^\mathrm{r} \geq 0 \\
			& 	D^{\mathrm{rp}} = \!\!\!\sum_{m\in \mathcal{M}}\!\!\Bigg(\sum_{\substack{n\in \mathcal{M}\\n>m}}\sum_{c\in \mathcal{P}_{mn}}\!\!\gamma_{mn}^c D^{mn,c}+ \gamma_{mm}D^{mm}\Bigg)\!\!\!\!\!\!\!\!\!\\
& \sum_{n\in \mathcal{M}\setminus \{m\}} \sum_{c\in \mathcal{P}_{mn}}\!\!\!\gamma_{mn}^c + 2 \gamma_{mm}  = \alpha_m, \forall m\in \mathcal{M}\\
		    & \gamma_{mn}^c \geq 0, \forall m,n\in \mathcal{M}:  n > m, \forall c\in \mathcal{P}_{mn}\\
		    & \gamma_{mm} \geq 0 , \forall m\in \mathcal{M} \\
		   	&x^\mathrm{p} \in \mathrm{TAP}(\mathcal{R}^\mathrm{p}, X,x^\mathrm{r}).
		\end{aligned}
	\end{equation*}
	\vspace{-0.5cm}
\end{prob}

With the inclusion of the BPR travel time function~\eqref{eq:BPR},  Problem~\ref{prob:ridescongestion} becomes non-convex. However, following the procedure described in~\cite{Wollenstein-BetechSalazarEtAl2021}, for $\rho =1$, the BPR can be approximated by a piece-wise linear function, which allows to cast Problem~\ref{prob:ridescongestion} as a QP, for given private vehicle flows $x^\mathrm{p}$. As a result, performing bi-level iterations of the QP and TAP until convergence provides an efficient solution to Problem~\ref{prob:ridescongestion}. Even though there are no convergence guarantees, empirically it always converges in few iterations and the solutions are consistent for different conditions. 


\begin{remark}
Notice that the computation of set of matrices $D^{mn,c}$ amounts to the serving three active flow requests between the origin and destination nodes of $r_m$ and $r_n$ according to the serving order of configuration $c$. As a result, these should be precomputed, as explained in Section~\ref{sec:SD}, before solving Problem~\ref{prob:ridescongestion}.
\end{remark}


\section{Case Study}\label{sec:res}
This section showcases the results of the proposed framework in a case study of Sioux Falls, USA and one of Manhattan, USA. The data for Sioux Falls is obtained from the Transportation Networks for Research repository~\cite{ResearchCoreTeam}. The network of Manhattan is based on~\cite{HaklayWeber2008}, while the travel requests are taken from~\cite{TaxiNYC}. 
The private travel requests $\mathcal{R}^\mathrm{p}$ and the AMoD requests $\mathcal{R}$ were obtained from the same travel demands by scaling the magnitude of users per unit time by $1-\phi$ and $\phi$, respectively. As a result, the overall number of travel requests remains constant independently of $\phi$. A MATLAB implementation is available in an open-source repository at {\small \url{https://github.com/fabiopaparella/congestionR_AMoD}}.

\subsection{Comparison of Congestion-aware and -unaware Ride-pooling Assignment}
\begin{figure}[t!]
	\centering
	\includegraphics[trim={15 0 20 10},clip,width=\linewidth]{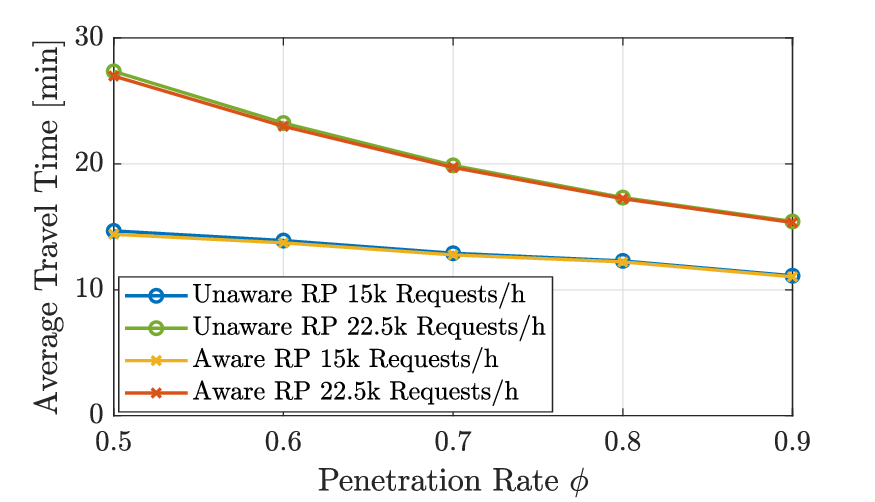}
	\caption{Comparison of average travel time experienced by AMoD users that ride-pool in Sioux Falls as a function of penetration rate and number of overall travel requests between a congestion-unaware ride-pooling assignment (Unaware RP) and a congestion-aware ride-pooling assignment (Aware RP), both with congestion-aware routing.}\label{fig:RPcomp}
	\vspace{-0.6cm}
\end{figure}

In this section, we compare the performance of the congestion-aware and -unaware ride-pooling assignment frameworks in terms of average user travel time and congestion on a link level in Sioux Falls. The congestion-aware assignment that was employed leverages bi-level QP and TAP iterations to solve Problem~\ref{prob:ridescongestion}. The solution converged within a relative variation of the objective of $10^{-2}$ always in less than $6$ iterations. Thereby, the congestion-unaware approach leverages the knapsack-like polynomial-time assignment scheme proposed in \cite{PaparellaPedrosoEtAl2023}, whereby the routing is based on Problem~\ref{prob:rides}. Fig.~\ref{fig:RPcomp} depicts the comparison of the average user travel time as a function of the penetration rate $\phi$  between a congestion-aware and a congestion-unaware ride-pooling assignment for two intensities of travel requests.
\begin{figure}[t!]	
	\centering
	\begin{subfigure}{\linewidth}
	\centering
	\begin{minipage}{0.44\linewidth}
		\includegraphics[trim={37 20 74 13},clip,width=0.99\linewidth]{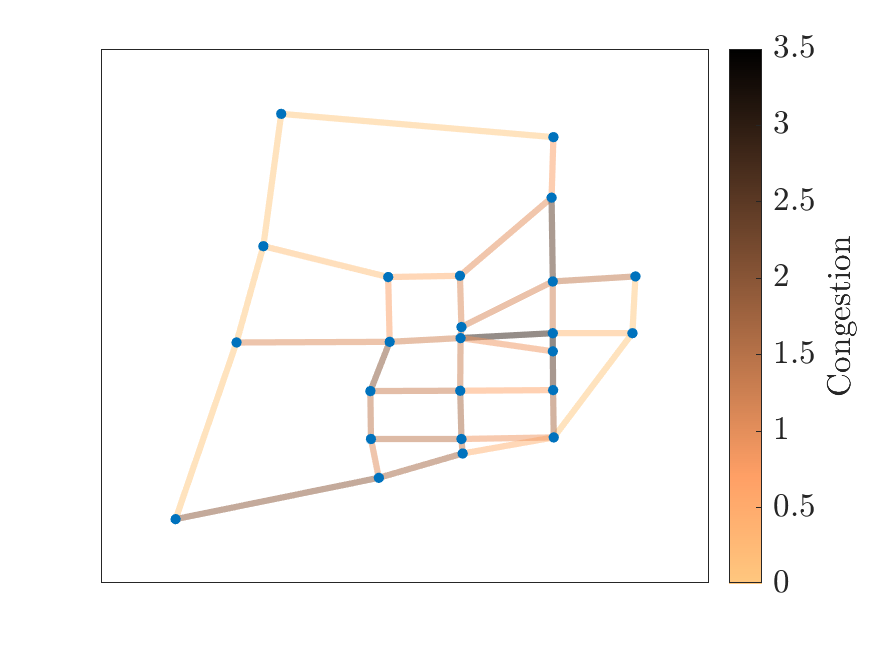}
		\end{minipage}
	\begin{minipage}{0.54\linewidth}
		\includegraphics[trim={36 20 5 13},clip,width=0.99\linewidth]{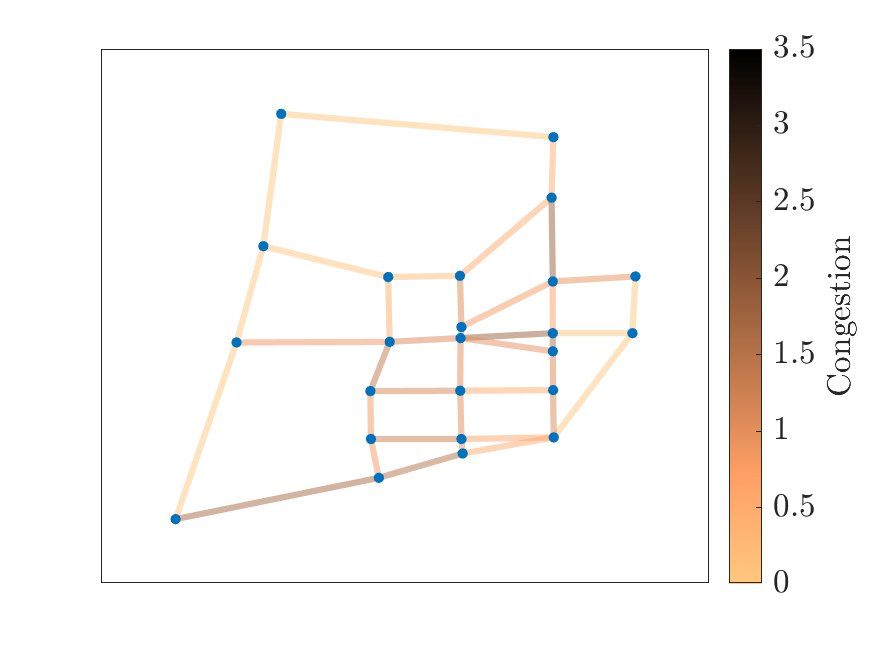}
		\end{minipage}
		\caption{Congestion level resulting from the congestion-unaware routing and assignment (left) and congestion-aware routing and assignment (right).}
		\label{fig:congestion_map}
	\end{subfigure}
	\begin{subfigure}{\linewidth}
	\centering
		\includegraphics[trim={10 0 10 0},clip,width=0.8\linewidth]{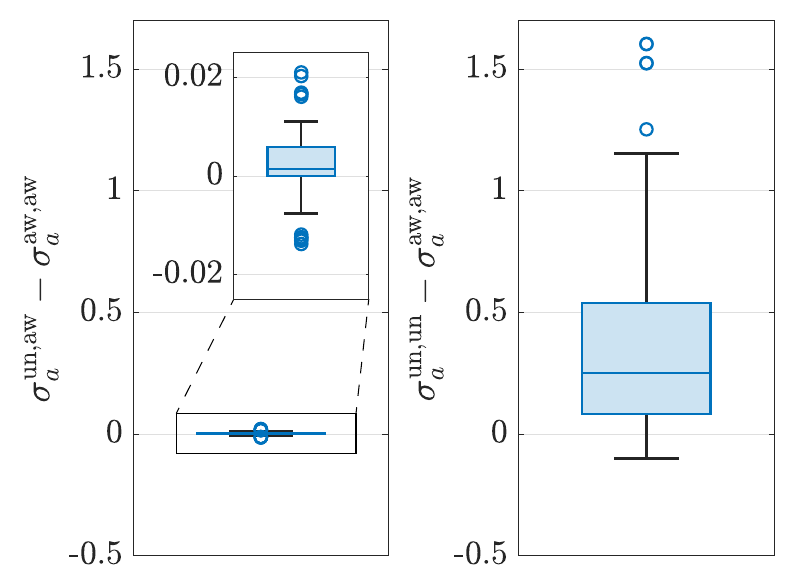}
		\caption{Distribution of the difference in congestion per link between i)~congestion-unaware assignment with \mbox{-aware} routing (left), and ii)~unaware assignment with congestion-unaware routing (right) and congestion-aware routing and assignment.}
		\label{fig:congestion_dif}
	\end{subfigure}
	\caption{Comparison of congestion levels in mixed traffic with \mbox{$\phi = 0.7$} between congestion-aware and -unaware assignment, and between aware and unaware routing for $30$ thousand travel requests per hour in Sioux Falls.}
	\label{fig:congestion}
	\vspace{-0.5cm}
\end{figure}%
We quantify congestion in a link $a$ with a flow $x_a$ and capacity $\kappa_a$  as $\sigma_a = \max(0,x_a - \kappa_a)/\kappa_a$, i.e., congestion is null in a link if the flow on that link is smaller or equal than capacity, and congestion equal to $\sigma_a>0$ if the flow is $(1+\sigma_a)$ times more than capacity.  Fig.~\ref{fig:congestion} shows the congestion on the links of the network for the congestion-aware ride-pooling assignment and routing with a penetration rate of $\phi = 0.7$ and the difference w.r.t.\ the congestion-unaware assignment scenario.

Interestingly, Fig.~\ref{fig:congestion_dif} shows that if the routing is congestion-aware, then the congestion patterns are practically the same. 
 On the contrary, we notice congestion-unaware routing leads to significant increase in congestion.
 To sum, we recognize that the ride-pooling matching congestion-awareness does not influence the congestion pattern, nor does it influence the average travel time, as long as the AMoD routing itself is congestion-aware. This shows that employing a congestion-unaware assignment (with congestion-aware routing) is a good compromise between complexity and tractability. In addition, the algorithm leveraged in congestion-unaware assignment could also be extended to take into account private flow congestion with $\phi=0$---i.e., neglecting endogenous effects---or be solved iteratively, similar to a fixed-point iteration, to also account for endogenous congestion.
 We leave this interesting aspect for future research. To sum up, the results show that even without any information about congestion, a congestion-unaware procedure is still a viable way to obtain a good ride-pooling assignment strategy.


 

\subsection{Effects of Penetration Rate}

\begin{figure*}[t!]
	\centering
	\begin{subfigure}{0.4\linewidth}
		\centering
		\includegraphics[trim={0 0 0 0},clip,width=.8\linewidth]{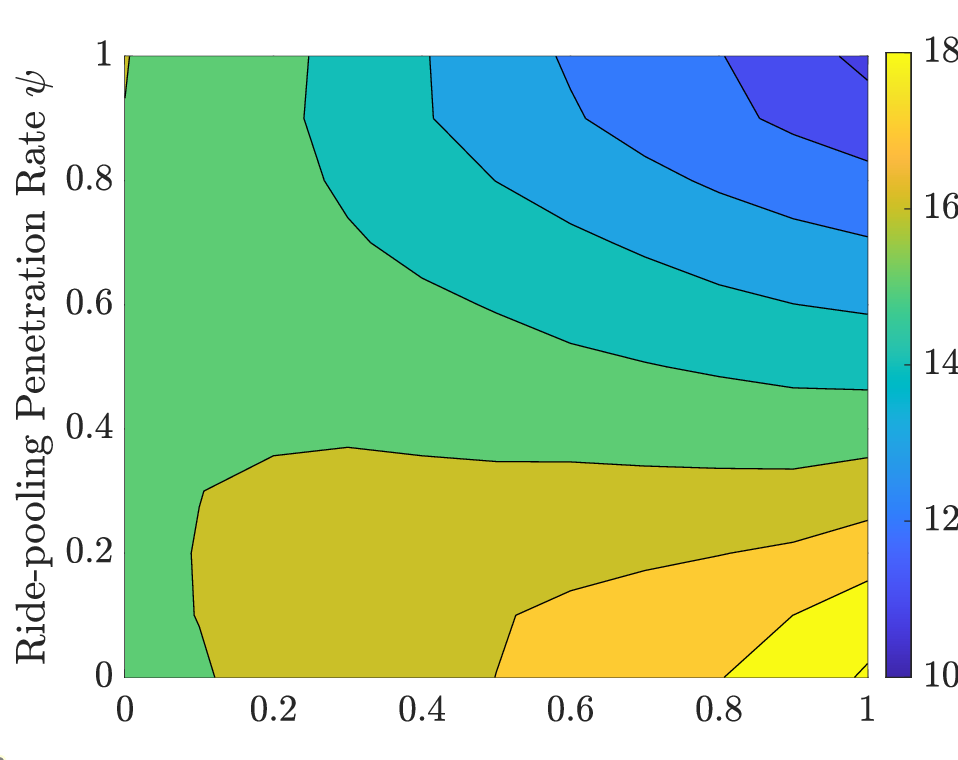}
		\caption{Individual ride-sharing average travel time.}
		\label{fig:phipsi_sharing}
	\end{subfigure}
	\hspace{.5cm}
	\begin{subfigure}{0.4\linewidth}
		\centering
		\includegraphics[trim={0 00 00 0},clip,width=.8\linewidth]{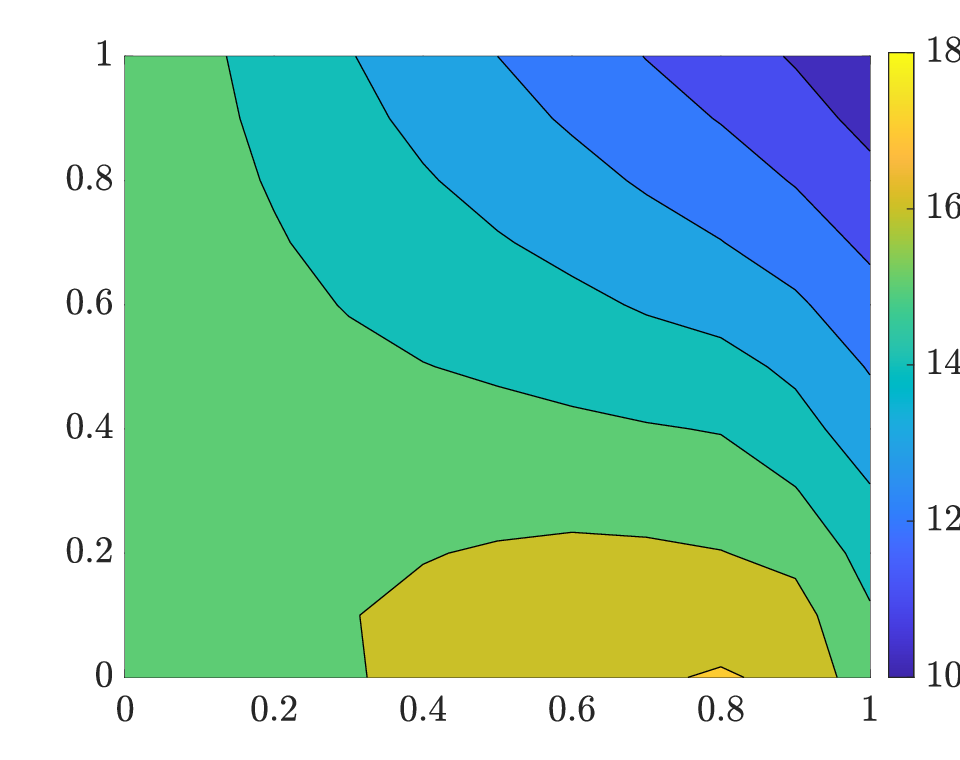}
		\caption{Private user average travel time.}
		\label{fig:phipsi_private}
	\end{subfigure}\\
	\begin{subfigure}{0.4\linewidth}
		\centering
		\includegraphics[trim={0cm 0 0 0},clip,width=.8\linewidth]{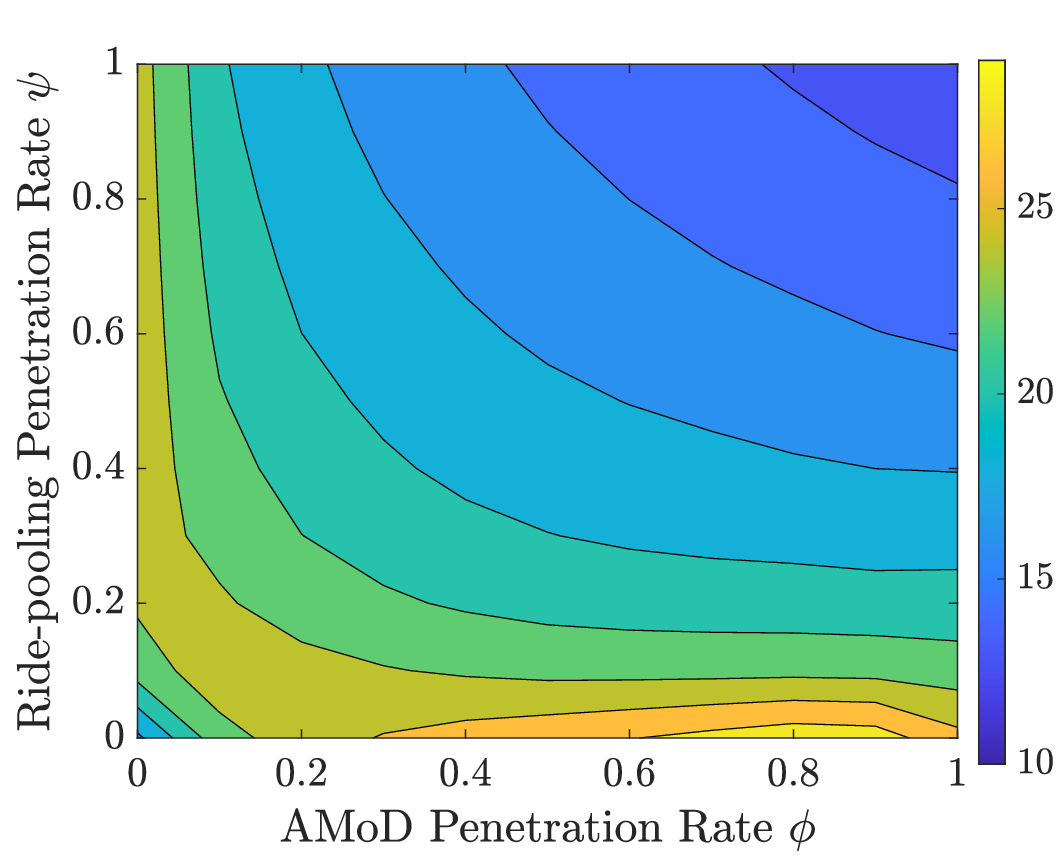}
		\caption{Ride-pooling average travel time.}
		\label{fig:phipsi_pooling}
	\end{subfigure}\hspace{.5cm}
	\begin{subfigure}{0.4\linewidth}
		\centering
		\includegraphics[trim={0 00 00 0},clip,width=.8\linewidth]{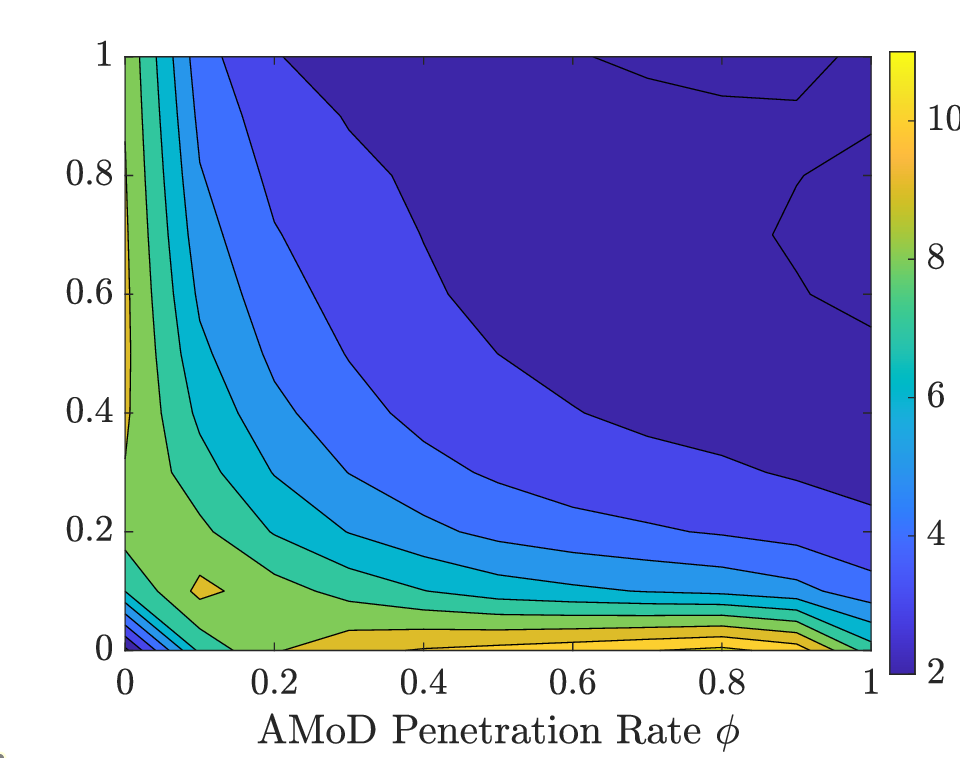}
		\caption{Difference between ride-pooling and individual ride-sharing.}
		\label{fig:phipsi_dif}
	\end{subfigure}
	\caption{Average travel time (in minutes) experienced by users in Manhattan as a function of AMoD penetration rate $\phi$ and ride-pooling penetration rate $\psi$, using the congestion-unaware ride-pooling assignment.}
	\label{fig:phipsi}
\end{figure*}

\begin{figure}[t!]
	\centering
	\includegraphics[trim={10 0 20 0},clip,width=\linewidth]{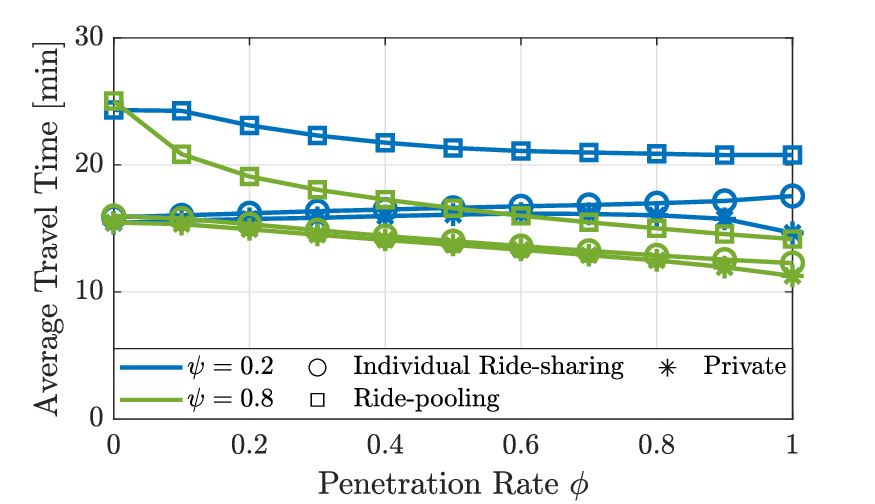}
	\includegraphics[trim={10 0 20 5},clip,width=\linewidth]{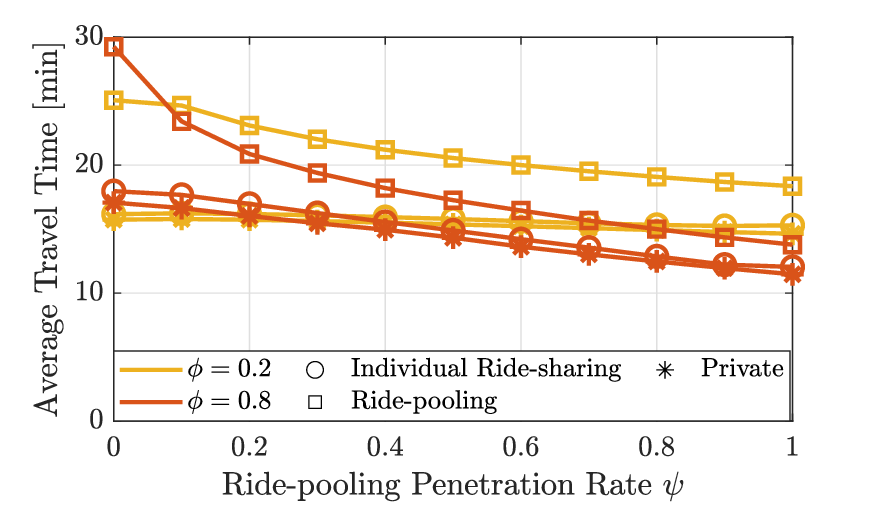}
	\caption{Average travel time experienced by private, ride-sharing and ride-pooling users in Manhattan as a function of AMoD penetration rate $\phi$ and ride-pooling penetration rate $\psi$, using the congestion-unaware ride-pooling assignment.}
	\label{fig:cut}
	\vspace{-0.1 cm}
\end{figure}

In this section, we showcase the results in Manhattan, where there are overall $212$ thousand travel requests divided into private users and AMoD users that can choose to either ride-pool or not. As before, $\phi$ denotes the fraction of AMoD requests w.r.t.\ the overall number of travel requests. We now denote $\psi$ as the penetration rate of ride-pooling requests withing the AMoD system, i.e., the fraction of ride-pooling travel requests w.r.t.\ the overall number of AMoD travel requests. It is important to remark that if a request that can be ride-pooled is assigned to be served individually without pooling it with any other request, it is still accounted for as a ride-pooling request.
Fig.~\ref{fig:phipsi} and Fig.~\ref{fig:cut} show the average travel time for the three types of users 
as a function of the AMoD penetration rate $\phi$ and the ride-pooling penetration rate $\psi$.

First, the results show that for every $\psi$ and $\phi$, the travel time of private users is always slightly below the individual ride-sharing travel-time, since the routing of private users is selfish and the routing of the AMoD fleet is system-optimal. Note that in this simulation the parking time of private users, which might degrade their combined travel time, is not modeled. Second, the average travel time of individual ride-sharing users is also always below the average ride-pooling travel time. Indeed, from Fig.~\ref{fig:phipsi_dif}, it amounts to a difference between $\unit[2]{min}$ and  $\unit[11]{min}$. Third, observing the average travel time of individual ride-sharing  and private users, in Figures~\ref{fig:phipsi_sharing} and \ref{fig:phipsi_private}, respectively, we notice that an increase of the penetration rate $\phi$, does not always translate into a decrease in travel time of individual ride-sharing users, even if they are centrally controlled. This behavior is due to the fact that, in contrast to private vehicles, the AMoD fleet has to be rebalanced, which additional empty-vehicle flows worsen congestion and, consequently, travel time. This is especially the case for a low ride-pooling penetration rate $\psi$, whereby very few users ride-pool. Indeed, for $\psi \approx 0.5$, the rebalancing-induced congestion is counter-balanced by the congestion reduction that stems from ride-pooling users. Moreover, from Fig.~\ref{fig:cut}, the average travel time of individual ride-sharing users increases with $\phi$ for $\psi = 0.2$ but decreases with $\phi$ for $\psi = 0.8$. Fourth, from Fig.~\ref{fig:phipsi_dif}, we notice that when there are few ride-pooling users, i.e., low $\psi$ and/or $\phi$, the additional average travel time of ride-pooling users w.r.t.\ individual ride-sharing is very high. On top of that, for a very low ride-pooling penetration rate $\psi$ and a high penetration rate $\phi$, due to both high congestion caused by rebalancing, and large detours caused by few ride-pooling users, the difference in average travel compared to individual ride-sharing users grows significantly. For $\psi,\phi \leq 0.1$, the number of ride-pooling users is so low that they cannot be ride-pooled due to time constraints, forcing the operator to abstain from pooling them with another user. As a result, although these users are available and willing to be ride-pooled, they do not and the difference in average travel time is low, as seen in Fig.~\ref{fig:phipsi_dif}. Fifth, as $\phi$ and $\psi$ grow, the number of ride-pooling users is high enough, so matching is possible, even if slightly inconvenient. For larger $\phi$ and $\psi$, ride-pooling becomes very efficient---in line the Mohring and Better matching effects~\cite{FielbaumTirachiniEtAl2021}---and can substantially decrease the detours and, in consequence, the congestion level and the travel time of all users, as seen in Fig.~\ref{fig:cut}. 

\section{Conclusions}\label{sec:conc}
This paper presented a framework to model ride-pooling Autonomous Mobility-on-Demand (AMoD) in a congestion-aware time-invariant network flow model. Specifically, by assuming there is a sufficiently high number of travel requests alongside other approximations, it is possible to frame the joint congestion-aware assignment and routing problem as a quadratic program that can be solved in polynomial-time.
First, the case study of Sioux Falls quantitatively showed that a congestion-aware ride-pooling assignment scheme does not have a significant impact on congestion or average travel time, as far as the routing is congestion-aware. As a result, we conclude that combining an congestion-unaware assignment scheme with congestion-aware routing is a good compromise between tractability and performance.
Second, the results of the case study of Manhattan revealed that ride-pooling can significantly contribute to lower congestion and travel times thanks to the lower number of vehicle hours traveled. What is more, individual ride-sharing alone, although routed in a system-optimal pattern, not only does not lower congestion but may also increase it due to rebalancing-induced congestion. Instead, for a high enough ride-pooling penetration rate, the increase in congestion caused by additional rebalancing trips of empty vehicles and detours to ride-pool requests is overcompensated by the significantly lower number of trips required to serve the requests, ultimately resulting in overall lower congestion and lower travel time for all users.

This work can be extended by integrating the mesoscopic ride-pooling model within intermodal settings~\cite{Wollenstein-BetechSalazarEtAl2021}, potentially also accounting for the energy consumption of the vehicles~\cite{PaparellaChauhanEtAl2023}.



\section*{Acknowledgments}\label{Sec:akn}
We thank Dr. I. New for proofreading the paper and L. Hollander for the support with Fig.~\ref{fig:eg_ridepooling}. This publication is part of the project NEON with number 17628 of the research program Crossover.

\bibliographystyle{IEEEtran}
\bibliography{main.bib,SML_papers.bib,./Sections/references_licensing.bib}

\newcommand{\noopsort}[1]{} \newcommand{\printfirst}[2]{#1}
  \newcommand{\singleletter}[1]{#1} \newcommand{\switchargs}[2]{#2#1}
\begin{thebibliography}{10}
\providecommand{\url}[1]{#1}
\csname url@samestyle\endcsname
\providecommand{\newblock}{\relax}
\providecommand{\bibinfo}[2]{#2}
\providecommand{\BIBentrySTDinterwordspacing}{\spaceskip=0pt\relax}
\providecommand{\BIBentryALTinterwordstretchfactor}{4}
\providecommand{\BIBentryALTinterwordspacing}{\spaceskip=\fontdimen2\font plus
\BIBentryALTinterwordstretchfactor\fontdimen3\font minus
  \fontdimen4\font\relax}
\providecommand{\BIBforeignlanguage}[2]{{%
\expandafter\ifx\csname l@#1\endcsname\relax
\typeout{** WARNING: IEEEtran.bst: No hyphenation pattern has been}%
\typeout{** loaded for the language `#1'. Using the pattern for}%
\typeout{** the default language instead.}%
\else
\language=\csname l@#1\endcsname
\fi
#2}}
\providecommand{\BIBdecl}{\relax}
\BIBdecl

\bibitem{Santi2014}
P.~Santi, G.~Resta, M.~Szell, S.~Sobolevsky, S.~H. Strogatz, and C.~Ratti,
  ``Quantifying the benefits of vehicle pooling with shareability networks,''
  \emph{{Proceedings of the National Academy of Sciences}}, vol. 111, no.~3,
  pp. 13\,290--13\,294, 2014.

\bibitem{Alonso_Mora_2017}
J.~Alonso-Mora, S.~Samaranayake, A.~Wallar, E.~Frazzoli, and D.~Rus,
  ``On-demand high-capacity ride-sharing via dynamic trip-vehicle assignment,''
  \emph{Proceedings of the National Academy of Sciences}, vol. 114, no.~3, pp.
  462--467, jan 2017.

\bibitem{ISC_license}
\BIBentryALTinterwordspacing
``{ISC} {License}.'' [Online]. Available:
  \url{https://github.com/lucide-icons/lucide/blob/main/LICENSE}
\BIBentrySTDinterwordspacing

\bibitem{Apache_license}
\BIBentryALTinterwordspacing
``Apache {License} {Version} 2.0.'' [Online]. Available:
  \url{https://www.apache.org/licenses/LICENSE-2.0.txt}
\BIBentrySTDinterwordspacing

\bibitem{TsaoMilojevicEtAl2019}
M.~Tsao, D.~Milojevic, C.~Ruch, M.~Salazar, E.~Frazzoli, and M.~Pavone, ``Model
  predictive control of ride-sharing autonomous mobility on demand systems,''
  in \emph{{Proc.\ IEEE Conf.\ on Robotics and Automation}}, 2019.

\bibitem{FielbaumKucharskiEtAl2022}
A.~Fielbaum, R.~Kucharski, O.~Cats, and J.~Alonso-Mora, ``How to split the
  costs and charge the travellers sharing a ride? aligning system’s optimum
  with users’ equilibrium,'' \emph{European Journal of Operational Research},
  vol. 301, no.~3, pp. 956--973, 2022.

\bibitem{FielbaumTirachiniEtAl2021}
A.~Fielbaum, A.~Tirachini, and J.~Alonso-Mora, ``Economies and diseconomies of
  scale in on-demand ridepooling systems,'' \emph{Economics of Transportation},
  vol.~34, p. 100313, 2023.

\bibitem{PavoneSmithEtAl2012}
M.~Pavone, S.~L. Smith, E.~Frazzoli, and D.~Rus, ``Robotic load balancing for
  {Mobility-on-Demand} systems,'' \emph{{Proc.\ of the Inst.\ of Mechanical
  Engineers, Part~D: Journal of Automobile Engineering}}, vol.~31, no.~7, pp.
  839--854, 2012.

\bibitem{ZhangPavone2014}
R.~Zhang and M.~Pavone, ``Control of robotic {Mobility-on-Demand} systems: a
  queueing-theoretical perspective,'' in \emph{{Robotics: Science and
  Systems}}, 2014, best Paper Award Finalist.

\bibitem{ZardiniLanzettiEtAl2022}
G.~Zardini, N.~Lanzetti, M.~Pavone, and E.~Frazzoli, ``Analysis and control of
  autonomous mobility-on-demand systems,'' \emph{{Annual Review of Control,
  Robotics, and Autonomous Systems}}, vol.~5, 2022.

\bibitem{Rossi2018}
F.~Rossi, ``On the interaction between {Autonomous Mobility-on-Demand} systems
  and the built environment: Models and large scale coordination algorithms,''
  Ph.D. dissertation, {Stanford University, Dept.\ of Aeronautics and
  Astronautics}, 2018.

\bibitem{RossiIglesiasEtAl2018b}
F.~Rossi, R.~Iglesias, M.~Alizadeh, and M.~Pavone, ``On the interaction between
  {Autonomous Mobility-on-Demand} systems and the power network: Models and
  coordination algorithms,'' \emph{{IEEE Transactions on Control of Network
  Systems}}, vol.~7, no.~1, pp. 384--397, 2020.

\bibitem{PaparellaPedrosoEtAl2023}
F.~Paparella, L.~Pedroso, T.~Hofman, and M.~Salazar, ``A time-invariant network
  flow model for two-person ride-pooling mobility-on-demand,'' in \emph{{Proc.\
  IEEE Conf.\ on Decision and Control}}, 2023, in press.

\bibitem{Wollenstein-BetechSalazarEtAl2021}
S.~Wollenstein-Betech, M.~Salazar, A.~Houshmand, M.~Pavone, C.~G. Cassandras,
  and I.~C. Paschalidis, ``Routing and rebalancing intermodal autonomous
  mobility-on-demand systems in mixed traffic,'' \emph{{IEEE Transactions on
  Intelligent Transportation Systems}}, vol.~23, no.~8, pp. 12\,263--12\,275,
  2021.

\bibitem{Bullo2018}
\BIBentryALTinterwordspacing
F.~Bullo, \emph{Lectures on Network Systems}, 1st~ed.\hskip 1em plus 0.5em
  minus 0.4em\relax Kindle Direct Publishing, 2020, with contributions by J.
  Cortes, F. Dorfler, and S. Martinez. [Online]. Available:
  \url{http://motion.me.ucsb.edu/book-lns}
\BIBentrySTDinterwordspacing

\bibitem{Patriksson2015}
M.~Patriksson, \emph{The traffic assignment problem: models and methods}.\hskip
  1em plus 0.5em minus 0.4em\relax Courier Dover Publications, 2015.

\bibitem{IglesiasRossiEtAl2017}
R.~Iglesias, F.~Rossi, R.~Zhang, and M.~Pavone, ``A {BCMP} network approach to
  modeling and controlling autonomous mobility-on-demand systems,''
  \emph{{Proc.\ of the Inst.\ of Mechanical Engineers, Part~D: Journal of
  Automobile Engineering}}, vol.~38, no. 2--3, pp. 357--374, 2019.

\bibitem{Neuburger1971}
H.~Neuburger, ``The economics of heavily congested roads,''
  \emph{{Transportation Research}}, vol.~5, no.~4, pp. 283--293, 1971.

\bibitem{PaparellaChauhanEtAl2023}
F.~Paparella, K.~Chauhan, T.~Hofman, and M.~Salazar, ``Electric autonomous
  mobility-on-demand: Joint optimization of routing and charging infrastructure
  siting,'' in \emph{{IFAC World Congress}}, 2023.

\bibitem{SalazarLanzettiEtAl2019}
M.~Salazar, N.~Lanzetti, F.~Rossi, M.~Schiffer, and M.~Pavone, ``Intermodal
  autonomous mobility-on-demand,'' \emph{{IEEE Transactions on Intelligent
  Transportation Systems}}, vol.~21, no.~9, pp. 3946--3960, 2020.

\bibitem{BPR1964}
{Bureau of Public Roads}, ``Traffic assignment manual,'' {U.S. Dept.\ of
  Commerce, Urban Planning Division}, Tech. Rep., 1964.

\bibitem{ResearchCoreTeam}
T.~N. for Research Core~Team. Transportation networks for research.
  https://github.com/bstabler/transportationnetworks. Accessed January, 05,
  2023.

\bibitem{HaklayWeber2008}
M.~Haklay and P.~Weber, ``{OpenStreetMap}: User-generated street maps,''
  \emph{{IEEE Pervasive Computing}}, vol.~7, no.~4, pp. 12--18, 2008.

\bibitem{TaxiNYC}
``Tlc trip record data,''
  \url{https://www.nyc.gov/site/tlc/about/tlc-trip-record-data.page}, accessed:
  11-2023.

\end{thebibliography}

\end{document}